\documentclass{ieeeaccess}
\usepackage{cite}
\usepackage{amsmath,amssymb,amsfonts}
\usepackage{algorithmic}
\usepackage[pdftex]{graphicx}
\usepackage{textcomp}
\def\BibTeX{{\rm B\kern-.05em{\sc i\kern-.025em b}\kern-.08em
    T\kern-.1667em\lower.7ex\hbox{E}\kern-.125emX}}
\usepackage{supertabular}
\usepackage{multirow}
\usepackage{threeparttable}
\newcommand{\bm}[1]{\mbox{\boldmath $#1$}}
\newfont{\bg}{cmr6 scaled\magstep4}
\newcommand{\bigzerol}{\smash{\lower0.6ex\hbox{\bg 0}}}

\begin{document}
\history{Date of publication xxxx 00, 0000, date of current version xxxx 00, 0000.}
\doi{10.1109/ACCESS.2017.DOI}

\title{A Neural Network Based on the Johnson $S_\mathrm{U}$ Translation System and Related Application to Electromyogram Classification}
\author{\uppercase{Hideaki Hayashi}\authorrefmark{1}, \IEEEmembership{Member, IEEE},
\uppercase{Taro Shibanoki}\authorrefmark{2}, \IEEEmembership{Member, IEEE},
\uppercase{and Toshio Tsuji}\authorrefmark{3}, \IEEEmembership{Member, IEEE}}
\address[1]{Department of Advanced Information Technology, Kyushu University, Fukuoka,  Japan (e-mail: hayashi@ait.kyushu-u.ac.jp)}
\address[2]{College of Engineering, Ibaraki University, Hitachi, Japan}
\address[3]{Department of System Cybernetics, Graduate School of Engineering, Hiroshima University, Higashi-hiroshima, Japan}
\tfootnote{This work was supported in part by JSPS KAKENHI Grant Number JP17K12752.}

\markboth
{Hayashi \headeretal: A Neural Network Based on the Johnson $S_\mathrm{U}$ Translation System and Related Application to Electromyogram Classification}
{Hayashi \headeretal: A Neural Network Based on the Johnson $S_\mathrm{U}$ Translation System and Related Application to Electromyogram Classification}

\corresp{Corresponding author: Hideaki Hayashi (e-mail: hayashi@ait.kyushu-u.ac.jp).}

\begin{abstract}
Electromyogram (EMG) classification is a key technique in EMG-based control systems. 
The existing EMG classification methods do not consider the characteristics of EMG features 
that the distribution has skewness and kurtosis, 
causing drawbacks such as the requirement of hyperparameter tuning. 
In this paper, we propose a neural network based on the Johnson $S_\mathrm{U}$ translation system 
that is capable of representing distributions with skewness and kurtosis.
The Johnson system is a normalizing translation 
that transforms non-normal data to a normal distribution, 
thereby enabling the representation of a wide range of distributions.
In this study, a discriminative model based on the multivariate Johnson $S_\mathrm{U}$ translation system 
is transformed into a linear combination of coefficients and input vectors using log-linearization. 
This is then incorporated into a neural network structure, 
thereby allowing the calculation of the posterior probability of the input vectors for each class and
the determination of model parameters as weight coefficients of the network. 
The uniqueness of convergence of the network learning is theoretically guaranteed.
In the experiments, the suitability of the proposed network for distributions including skewness and kurtosis 
is evaluated using artificially generated data. 
Its applicability for real biological data is also evaluated via an EMG classification experiment. 
The results show that the proposed network achieves high classification performance without the need for hyperparameter optimization. 
\end{abstract}


\titlepgskip=-15pt

\maketitle

{\allowdisplaybreaks
	\section{Introduction}
	Biosignals such as electroencephalograms (EEGs), electrocardiograms \allowbreak (ECGs), and electromyograms (EMGs) strongly reflect a human's internal state and intentions, 
and have therefore been applied to human--machine interfaces and diagnosis \cite{mammone2018permutation, sakhavi2018learning, zeng2016optimizing, zhang2016sparse}. 
In particular, EMG-based control systems have been widely studied, because EMGs can be voluntarily controlled. 
Many practical applications have been developed, typified by myoelectric prosthetics, which are prosthetic hands that can be controlled using surface EMGs \cite{tigra2016novel, fukuda2003human}. 

According to Oskoei and Hu \cite{asghari2007myoelectric}, EMG-based control systems include four main stages: 
data segmentation, feature extraction, classification, and control. 
The raw data are first segmented, and then converted into feature vectors. 
These vectors are classified into predefined categories, 
before the controller generates output commands for the instruments based on the classification results. 

To realize highly intuitive and dexterous control, 
it is particularly important to achieve a high level of classification performance, both in terms of accuracy and speed of training and prediction. 
Classifiers such as the support vector machine (SVM) \cite{cortes1995support}, 
multilayer perceptron (MLP) with back propagation learning \cite{rummelhart1986learning}, 
and $k$-nearest neighbors algorithm ($k$-NN) \cite{cover1967nearest} have been widely used. 
These popular techniques, however, are not always the most suitable for EMG classification 
in spite of their high classification abilities. 
For example, SVMs are computationally expensive for hyperparameter optimization, and
the MLP requires a very long training time. Similarly, it is difficult to use $k$-NN in real-time applications because of the large computational cost of prediction.

To improve the classification ability for a certain purpose, 
stochastic models can be incorporated into the structure of the classifier 
if there is some prior knowledge of the input signals \cite{wang2017g, hayashi2015recurrent, tsuji2003recurrent, tsuji1999log}. 
For instance, Tsuji {\it et al.} \cite{tsuji1999log} proposed a Gaussian mixture model-based neural network, known as the log-linearized Gaussian mixture network (LLGMN), 
by assuming that the input signals obey a Gaussian mixture model. 

The authors now consider the following assumptions to be the {\it prior knowledge} of the feature vectors obtained from EMG signals: 
\begin{itemize}
\item The distribution of input signals for each class is unimodal
\item The distribution has skewness and kurtosis
\end{itemize}
The derivation of these assumptions are explained in Section III.

To satisfy the above assumptions, 
we can use a flexible distribution known as the Johnson distribution \cite{johnson1949systems}, 
which represents the mean, variance, skewness, and kurtosis using four parameters. 
Its extension to higher dimensions is enabled by the multivariate Johnson translation system \cite{stanfield1996multivariate,johnson1949bivariate,johnson1972distributions}.
If we could construct a classifier that incorporates the multivariate Johnson distribution in its structure, 
it would be applicable to EMG classification and EMG-based systems. 

This paper proposes a neural network (NN) based on the Johnson $S_\mathrm{U}$ translation system. 
The proposed NN represents a flexible distribution by including a discriminative model based on the multivariate Johnson $S_\mathrm{U}$ translation system, 
thereby supporting the accurate classification of data with skewness and kurtosis. 
The parameters of the model can be determined as weight coefficients of the proposed NN via learning. 

This paper is related to our previous workshop paper \cite{hayashi2015non}. 
Our previous work was preliminary, and has the following drawbacks: 
\begin{itemize}
\item The network structure does not correctly represent the Johnson distribution due to the lack of the Jacobian.
\item The training algorithm does not guarantee the uniqueness of convergence. 
\item The dataset variation and comparison are limited in the EMG classification experiments.
\end{itemize}
In this paper, the above problems have been solved. 

The rest of this paper is organized as follows: 
Related studies and their characteristic comparisons are described in Section II.
Section III explains the derivation of the above assumptions, 
then describes a discriminative model based on the multivariate Johnson translation system 
and its transformation to linear combinations of weight coefficients and input vectors via log-linearization. 
The structure and learning algorithms of the proposed NN are presented in Section IV, 
and the results of a simulation experiment using artificial data are described in Section V. 
Section VI outlines the application potential for biosignal classification based on an EMG classification experiment. 
Finally, Section VII concludes the paper.

	\section{Related work}
	This section summarizes popular algorithms for EMG classification, 
and compares each of their characteristics. 
The algorithms compared in this section are: 
\begin{itemize}
\item SVM \cite{cortes1995support}
\item LLGMN \cite{tsuji1999log}
\item MLP \cite{rummelhart1986learning}
\item Linear logistic regression (LLR) \cite{bishop2006pattern}
\item $k$-NN \cite{cover1967nearest}
\item Random forest
\end{itemize}

These algorithms are compared in terms of the following significant factors for EMG classification and EMG-based control systems: 
(non-)requirement of hyperparameter optimization, speed of training, uniqueness of solutions, speed of prediction, nonlinearity, and computability of posterior probabilities. 
The first three factors are associated with the effort needed to construct the classifier. 
Hyperparameter optimization is conducted before training, and typically takes a very long time; hence, the usability of the system is enhanced if this step is not required. 
Fast training and a unique solution are also desirable to avoid effort and uncertainty in the training of the classifier. 
Systems that are to be deployed in an online manner require fast prediction, while
the nonlinearity and computability of posterior probabilities is related to the accuracy of classification. 
Although EMG classification problems are unlikely to be linear, 
their nonlinearity is not overly complex, because each class of EMG signals can be clustered to some extent. 
Calculating the posterior probabilities has some powerful merits, such as minimizing risk, rejecting options, compensating for class priors, and combining models \cite{bishop2006pattern}. 

Table \ref{characteristics} summarizes the characteristics of the classification algorithms. 
\begin{table*}[t]
\centering
\caption{Characteristics of classification algorithms}
\begin{tabular}{lllllll} \hline
Algorithm 	& Hyperparameter-free	& Fast training  	& Unique solution 	& Fast prediction 	& Nonlinearity 	& Posterior probability	\\ \hline
SVM 		& $\times$ 						& \checkmark 		& \checkmark 		& $\times$ 			& \checkmark 	& $\times$				\\
LLGMN 		& $\times$ 						& $\times$ 			& $\times$ 			& \checkmark 		& \checkmark 	& \checkmark			\\
MLP 		& $\times$ 						& $\times$ 			& $\times$ 			& \checkmark 		& \checkmark 	& $\times$				\\
LLR 		& \checkmark 					& \checkmark 		& \checkmark 		& \checkmark 		& $\times$ 		& \checkmark			\\
$k$-NN 		& $\times$ 						& n/a 				& n/a 				& $\times$ 			& \checkmark 	& $\times$				\\ \hline
\end{tabular}
\label{characteristics}
\end{table*}
An SVM is distinguished classifier that realizes fast training and a unique solution. 
Its problem, however, is that two hyperparameters must be optimized. 
Additionally, because an SVM was originally developed as binary classifier,
multi-class classification can take a long time. 

The LLGMN is a discriminative model that incorporates Gaussian mixture models into an NN structure, 
allowing the posterior probability to be accurately calculated. 
The number of components (how many Gaussian distributions are summed in the model) should be carefully determined,
because the classification ability of the LLGMN for data following a non-Gaussian distribution decreases when there are few components.

The MLP is generally more compact than an SVM, and hence gives faster predictions, 
although training has a large computational cost. 
The number of layers and units should be determined as hyperparameters. 

The LLR is a probabilistic discriminative model that can be trained using Newton's method. 
The structural limit of the LLR is its inability to solve nonlinear separation problems. 

The $k$-NN is a very simple algorithm that does not require training. 
However, predictions entail large computational expense, as $k$-NN compares the distance between the input vector and every training vector. 
The constant $k$, the number of nearest neighbors used in the voting, is a user-defined constant, 
and is selected by heuristic techniques in general.

The proposed NN is designed to optimize the above characteristics as much as possible. 
In particular, there is no need to optimize the hyperparameters, and the uniqueness of solutions and computability of posterior probabilities can be theoretically explained.

	\section{Model structure}
	\subsection{Skewness and kurtosis in processed EMG signals}
After EMG signals have been acquired, we extract significant features. 
Although the raw EMG signals can be considered to obey a Gaussian distribution with zero mean \cite{hogan1980myoelectric}, 
the distribution of the extracted features may exhibit some skewness and kurtosis. 
We describe how the features obtain skewness and kurtosis in the process of feature extraction. 
Although many feature extraction methods have been proposed \cite{khushaba2017framework, nazmi2016review, altin2016comparison}, 
we focus on the method of Fukuda {\it et al.} \cite{fukuda2003human} because of its simplicity and universality.

Fukuda's method consists of two main parts: rectification and smoothing based on a Butterworth low-pass filter. 
Rectification takes the absolute value of the raw EMG signals, converting negative EMG values to positive values. 
Rectification is also used in methods such as 
integrated EMG (IEMG), mean absolute value (MAV), and modified mean absolute value (MMAV) \cite{rechy2011stages}, 
and is strongly related to the occurrence of skewness and kurtosis. 
Let $x$ be a raw EMG signal that obeys a Gaussian distribution with a mean of 0 and a standard deviation of $\sigma$. 
The skewness and kurtosis of $x$ are both 0.
Fig. \ref{Histogram} (a) shows an example raw EMG signal $x$ and its histogram.

\Figure[!t]()[width=0.99\hsize]{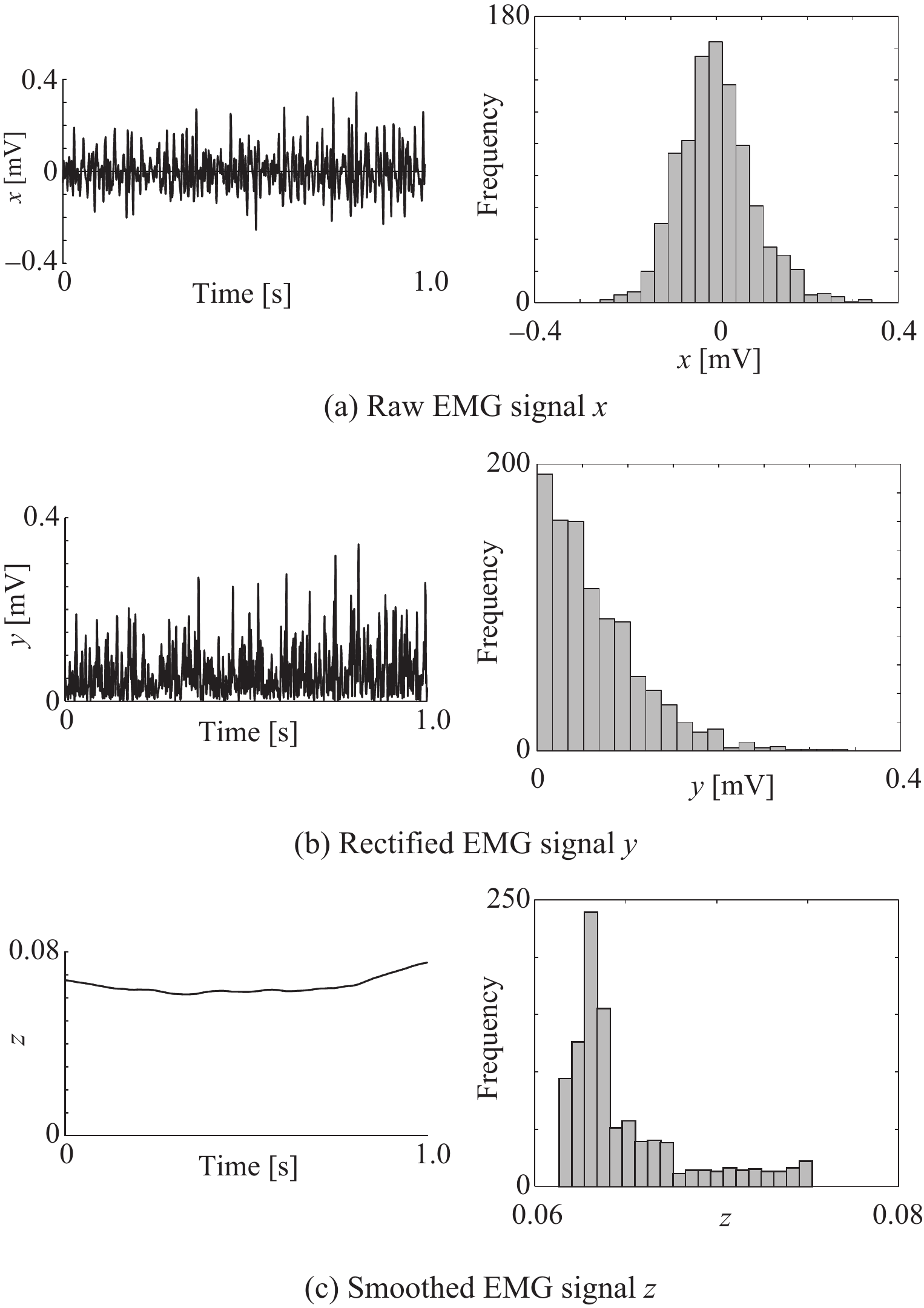}{Examples of the time-series signal and histogram of (a) raw EMG $x$, (b) rectified EMG $y$, and (c) smoothed EMG $z$. (a) raw EMG $x$ obeys a Gaussian distribution with zero mean (this is also discussed in Hogan and Mann \cite{hogan1980myoelectric}). The histograms of (b) rectified EMG $y$ and (c) smoothed EMG $z$, however, become asymmetric and include skewness and kurtosis.\label{Histogram}}
The probability density function of the rectified EMG signal $y = |x|$ is represented as
\begin{IEEEeqnarray}{rCl}
p(y) &=& \left\{
\begin{array}{ll}
\frac{2}{\sqrt{2\pi \sigma^2}}\exp \left[ -\frac{y^2}{2\sigma^2}\right] & (0 \leq y < \infty) \\
0 & (-\infty < y < 0)
\end{array}
\right..
\end{IEEEeqnarray}
The mean $M_y$ and the variance $V_y$ of $y$ is then calculated as 
\begin{IEEEeqnarray}{rCl}
M_y &=& \int_{-\infty}^{\infty}yp(y)dy = \sqrt{\frac{2}{\pi}}\sigma, \\
V_y &=& \int_{-\infty}^{\infty}(y-M_y)^2p(y)dy = (1-\frac{2}{\pi})\sigma^2.
\end{IEEEeqnarray}
In the rectified signal $y$, the skewness $S_y$ and the kurtosis $K_y$ are no longer $0$. They can be calculated as follows \cite{kim2004more}: 
\begin{IEEEeqnarray}{rCl}
S_y &=& \frac{\int_{-\infty}^{\infty}(y-M_y)^3 p(y)dy}{V_y^{\frac{3}{2}}} \!=\! \frac{(\frac{4}{\pi}-1)\sqrt{\frac{2}{\pi}}}{(1-\frac{2}{\pi})^\frac{3}{2}} \neq 0, \\
K_y &=& \frac{\int_{-\infty}^{\infty}(y-M_y)^4 p(y)dy}{V_y^2}-3 \nonumber \\
	&=& \frac{(3-\frac{4}{\pi}-\frac{12}{\pi^2})}{(1-\frac{2}{\pi})^2}-3 \neq 0.
\end{IEEEeqnarray}
The influence of rectification is also visible in the smoothed signal $z$ (see Fig. \ref{Histogram} (c)). 

Because the extracted features include skewness and kurtosis, 
conventional Gaussian-based models cannot readily model the data. 
In the next subsection, we therefore adopt the multivariate Johnson translation system \cite{stanfield1996multivariate}, which is suitable for data with skewness and kurtosis.

\subsection{Multivariate Johnson translation system}
The Jonson translation system is a family of transformations of a non-normal variate to a standard normal variate proposed by N. L. Johnson in 1949 \cite{johnson1949systems}. 
Based on this translation, Johnson derived a system of distributions that is suitable for representing distributions with skewness and kurtosis. 
Its multivariate extension is also proposed \cite{stanfield1996multivariate}.

Consider a $d$-dimensional continuous random vector $\bm{x} \in \mathbb{R}^{d}$ 
with skewness and kurtosis.
The multivariate Johnson translation system \cite{stanfield1996multivariate} involves the normalizing translation: 
\begin{equation}
\label{JohnsonTrans}
\bm{z} = \bm{\gamma} + \bm{\delta}\bm{g}[{\bm{\lambda}}^{-1}(\bm{x}-\bm{\xi})] \sim \mathcal{N}(\bm{0},\bf{\Sigma}),
\end{equation}
where $\bm{z}$ is a random vector obeying a normal distribution with mean $\bf{0}$ and variance $\bf{\Sigma}$, 
$\bm{\gamma} \equiv {[\gamma_1, \ldots, \gamma_d]}^\mathrm{T}$ and $\bm{\delta} \equiv \mathrm{diag}[\delta_1, \ldots, \delta_d]$ 
are shape parameters, $\bm{\lambda} \equiv \mathrm{diag}[\lambda_1, \ldots, \lambda_d]$ is a scale parameter, 
$\bm{\xi} \equiv {[\xi_1, \ldots, \xi_d]}^\mathrm{T}$ is a location parameter, 
and $\bm{g}(\cdot) \equiv {[g_1(\cdot), \ldots, g_d(\cdot)]}^\mathrm{T}$ denotes the transformation function that determines the family of a system.
$g_i(\cdot)$ $(i = 1, \ldots, d)$ is defined by the following four functions: 
\begin{equation}
g_i(y) = \left\{ 
\begin{array}{ll}
\!\ln{(y)} \!&\! \mathrm{for}~S_\mathrm{L}~(\mathrm{lognormal})\\
\!\ln \left[y+\sqrt{y^2+1}\right] \!&\! \mathrm{for}~S_\mathrm{U}~(\mathrm{unbounded})\\
\!\ln \left[y/(1-y)\right] \!&\! \mathrm{for}~S_\mathrm{B}~(\mathrm{bounded})\\
\!y \!&\! \mathrm{for}~S_\mathrm{N}~(\mathrm{normal})\\
\end{array}
\right.. 
\end{equation}
The domains of $x_i$ for $S_\mathrm{L}$, $S_\mathrm{U}$, $S_\mathrm{B}$, and $S_\mathrm{N}$ are 
$(\xi, +\infty)$, $(-\infty, +\infty)$, $(\xi, \xi+\lambda)$, and $(-\infty, +\infty)$, respectively. 
In (\ref{JohnsonTrans}), parameters $\bm{\lambda}$ and $\bm{\xi}$ affects the location and scale of the distribution of $\bm{x}$, respectively.
The combination of $\bm{\gamma}$ and $\bm{\delta}$ are associated with skewness and kurtosis,  
and $\bm{g}[\cdot]$ decides the shape of distribution tails, i.e. whether the distribution tails have boundary or go to infinity.

Since EMG can be seen as a random process, 
the systems with bounds are not suitable for EMG classification 
because probabilities cannot be calculated if an observation is out of the domain.
In unbounded systems, $S_\mathrm{U}$ is expected to be an extension of the normal distribution in particular, 
and have enough flexibility to fit data from arbitrary unimodal distribution.
This paper therefore focuses on $S_\mathrm{U}$ as the form of the function $g_i(y)$. 
Fig. \ref{JohnsonSUdist} shows an example of the Johnson $S_\mathrm{U}$ distribution, 
which can be calculated from the Johnson $S_\mathrm{U}$ translation system ($d = 1$).
\Figure[t]()[width=0.95\hsize]{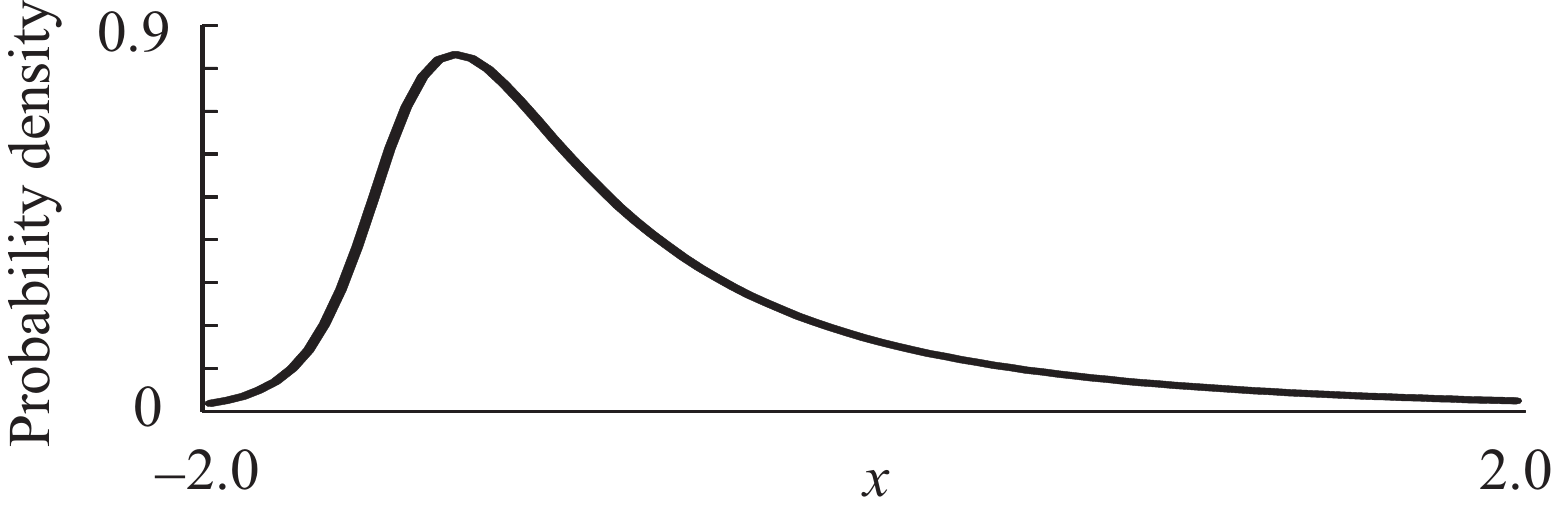}{Probability density function of the Johnson SU distribution ($d = 1$, $\gamma_1 = -1.4$, $\delta_1 = 1.0$, $\lambda_1 = 0.3$, $\xi_1 = -1.5$). This distribution represents skewness and kurtosis, hence the shape becomes asymmetric.\label{JohnsonSUdist}}
This asymmetric distribution represents skewness and kurtosis, 
and seems to be adaptable to the histogram of smoothed EMG shown in Fig. \ref{Histogram} (c).

\subsection{Posterior probability estimation}
To classify the vector $\bm{x} \in \mathbb{R}^{d}$ into one of the given $C$ classes, 
we must examine the posterior probability $P(c|\bm{x})$ $(c = 1, \ldots, C)$.
First, $\bm{x}$ is translated into a vector $\bm{z}^{(c)}$
using (\ref{JohnsonTrans}), which has the four parameters $\bm{\gamma}^{(c)}$, $\bm{\delta}^{(c)}$, $\bm{\lambda}^{(c)}$, and $\bm{\xi}^{(c)}$.

Assuming that the translated vector obeys a normal distribution, 
the posterior probability of $\bm{x}$ for class $c$ is calculated as
\begin{IEEEeqnarray}{rCl}
\label{PosteriorProb}
P(c|\bm{x}) &=& \frac{P(c)P(\bm{x}|c)}{\sum^{C}_{c=1}P(c)P(\bm{x}|c)},\\
\label{NormalDist}
P(\bm{x}|c) &=& \frac{|\mathbf{J}^{(c)}|}{(2\pi)^{\frac{d}{2}}{|\bf{\Sigma}^{(c)}|}^{\frac{1}{2}}}
\exp{\left(-\frac{1}{2}{\bm{z}^{(c)}}^\mathrm{T}{\bf{\Sigma}^{(c)}}^{-1}\bm{z}^{(c)}\right)},
\end{IEEEeqnarray}
where $P(c)$ is the prior probability of $c$, $\bf{\Sigma}^{(c)}$ is the variance matrix of $\bm{z}^{(c)}$, 
and $\mathbf{J}^{(c)}$ is the $d \times d$ Jacobian matrix, whose $(i, j)$th element is given by 
\begin{IEEEeqnarray}{rCl}
\frac{\partial z^{(c)}_i}{\partial x_j} \!=\! \left\{
\begin{array}{ll}
\delta_i^{(c)}{\lambda_i^{(c)}}^{-1}\!g'[{\lambda_i^{(c)}}^{-1}(x_i-\xi_i^{(c)})]  \!\!&\!\! (i \!=\! j) \\
0 \!\!&\!\! (i \!\neq\! j)
\end{array}
\!\!\right.,
\end{IEEEeqnarray}
where
\begin{IEEEeqnarray}{rCl}
g'_i(y) = \left\{ 
\begin{array}{ll}
\! 1/y \!&\! \mathrm{for}~S_\mathrm{L}~(\mathrm{lognormal})\\
\! 1/\sqrt{y^2+1} \!&\! \mathrm{for}~S_\mathrm{U}~(\mathrm{unbounded})\\
\! 1/[y(1-y)] \!&\! \mathrm{for}~S_\mathrm{B}~(\mathrm{bounded})\\
\! 1 \!&\! \mathrm{for}~S_\mathrm{N}~(\mathrm{normal})\\
\end{array}
\right..
\end{IEEEeqnarray}
The determinant of $\mathbf{J}^{(c)}$ is therefore calculated as
\begin{IEEEeqnarray}{rCl}
|\mathbf{J}^{(c)}| \!=\! \prod_{i=1}^d \frac{\partial z^{(c)}_i}{\partial x_i}
\!=\! \prod_{i=1}^d \left\{\frac{\delta_i^{(c)}}{\lambda_i^{(c)}}g'\left[\frac{(x_i-\xi_i^{(c)})}{\lambda_i^{(c)}}\right]\right\}.
\label{detJacobain}
\end{IEEEeqnarray}

\subsection{Log-linearization}
To incorporate the probabilistic model described above into a network structure, 
we transform the calculation of the Johnson translation and posterior probability estimation to linear combinations of 
coefficient matrices and input vectors. 

First, let $\bm{y}^{(c)}$ be a calculation in the function $\bm{g}(\cdot)$ of (\ref{JohnsonTrans}). 
$\bm{y}^{(c)}$ is then transformed as follows: 
\begin{IEEEeqnarray}{rCl}
\bm{y}^{(c)} &=& {\bm{\lambda}^{(c)}}^{-1}(\bm{x}-\bm{\xi}^{(c)})\nonumber\\
		 &=& {\bm{\lambda}^{(c)}}^{-1}\bm{x}-{\bm{\lambda}^{(c)}}^{-1}\bm{\xi}^{(c)}\nonumber\\
		 &=& \left[\!\!
\begin{array}{cccc}
-{\lambda^{(c)}}^{-1}_1\xi_{1}^{(c)} & {\lambda_{1}^{(c)}}^{-1} & & \bigzerol \\
\vdots & & \ddots & \\
-{\lambda^{(c)}}^{-1}_d\xi_{d}^{(c)} & \bigzerol & &{\lambda_{d}^{(c)}}^{-1}\\
\end{array} 
\!\!\right]\!
\left[\!\!
\begin{array}{c}
1 \\
\bm{x}
\end{array} 
\!\!\right] \nonumber\\
&=& {{}^{(1)}\mathbf{W}^{(c)}}^\mathrm{T}\mathbf{X}.
\label{y_trans}
\end{IEEEeqnarray}
Hence, $\bm{y}^{(c)}$ is expressed by multiplying the coefficient matrix ${}^{(1)}\mathbf{W}^{(c)} \in$ \allowbreak $\mathbb{R}^{(d+1) \times d}$ 
and the augmented input vector $\mathbf{X} \in \mathbb{R}^{d+1}$ .

Second, the translated vector $\bm{z}^{(c)}$ is also transformed and expressed as the product of 
a coefficient matrix and an augmented vector as follows: 
\begin{IEEEeqnarray}{rCl}
\bm{z}^{(c)} &=& \bm{\gamma}^{(c)} + \bm{\delta}^{(c)}\bm{g}\left(\bm{y}^{(c)}\right)\nonumber\\
		 &=& \left[\!\!
\begin{array}{cccc}
\gamma^{(c)}_1 & \delta^{(c)}_{1} & & \bigzerol \\
\vdots & & \ddots & \\
\gamma^{(c)}_d & \bigzerol & &\delta^{(c)}_{d}\\
\end{array} 
\!\!\right]\!
\left[\!\!
\begin{array}{c}
1 \\
\bm{g}\left(\bm{y}^{(c)}\right)
\end{array} 
\!\!\right] \nonumber\\
&=& {{}^{(2)}\mathbf{W}^{(c)}}^\mathrm{T}\mathbf{Y}^{(c)},
\label{z_trans}
\end{IEEEeqnarray}
where ${}^{(2)}\mathbf{W}^{(c)} \in \mathbb{R}^{(d+1) \times d}$ is a coefficient matrix 
and $\mathbf{Y}^{(c)} \in \mathbb{R}^{d+1}$ is determined by a nonlinear transformation of $\bm{y}^{(c)}$.

Finally, setting 
\begin{equation}
\zeta_{c} = P(c)P(\bm{x}|c)
\end{equation}
and taking the log-linearization of $\zeta_{c}$ gives
\begin{IEEEeqnarray}{rCl}
\log\zeta_{c} &=& [\log{P(c)} \!+\! \sum^d_{i=1}\log\frac{\delta_i^{(c)}}{\lambda_i^{(c)}} \!-\! \frac{1}{2}\log{|\bf{\Sigma}^{(c)}|}, -\frac{1}{2}{s}_{1,1}^{(c)},\nonumber \\
& & -{s}_{1,2}^{(c)}, \ldots, -\frac{1}{2}(2 - {\delta}_{i,j}){s}_{i,j}^{(c)}, \ldots\! ,\!-\frac{1}{2}{s}_{d,d}^{(c)}]\mathbf{Z}^{(c)} \nonumber \\
& & \!+\! \log \prod_{i=1}^d g'(y_i^{(c)})\nonumber \\
&\!\!\!=\!\!\!& {{}^{(3)}\mathbf{W}^{(c)}}^\mathrm{T}\mathbf{Z}^{(c)} \!+\! \sum_{i=1}^d \log g'(y_i^{(c)}),
\label{loglinearization}
\end{IEEEeqnarray}
where ${s}_{1,1}^{(c)}, \ldots, {s}_{d,d}^{(c)}$ are elements of the inverse matrix ${\bf{\Sigma}^{(c)}}^{-1}$, 
and ${\delta}_{i,j}$ is the Kronecker delta (1 if $i=j$, 0 otherwise).
Note that $(2\pi)^{-\frac{d}{2}}$ in (\ref{NormalDist}) is omitted because it is canceled out in (\ref{PosteriorProb}).
Additionally, $\mathbf{Z}^{(c)} \in \mathbb{R}^{H}$ $(H = 1+\frac{d(d+1)}{2})$ is defined as 
\begin{IEEEeqnarray}{rCl}
\label{nonlinearTrans}
\mathbf{Z}^{(c)}&=& [1, {z^{(c)}_{1}}^2, z^{(c)}_{1}z^{(c)}_{2}, \ldots, z^{(c)}_{1}z^{(c)}_{d}, \nonumber \\
& &{z^{(c)}_{2}}^2, z^{(c)}_{2}z^{(c)}_{3}, \ldots, {z^{(c)}_{d}}^2].
\end{IEEEeqnarray}
Taking the exponent of (\ref{loglinearization}), $\zeta_{c}$ (i.e., the numerator of (\ref{PosteriorProb})) is ultimately expressed by
\begin{equation}
\zeta_{c} = \exp\left({{}^{(3)}\mathbf{W}^{(c)}}^\mathrm{T}\mathbf{Z}^{(c)} \!+\! \sum_{i=1}^d \log g'(y_i^{(c)})\right)
\end{equation}

As outlined above, the Johnson translation and posterior probability estimation are calculated as linear combinations of coefficient matrices and nonlinearly transformed input vectors.
If these coefficients are appropriately determined, 
the parameters and structure of the model can be defined, 
and therefore the posterior probability of the input vectors can be calculated for each class.

The next section describes how the NN weight coefficients ${}^{(1)}\mathbf{W}^{(c)}$, ${}^{(2)}\mathbf{W}^{(c)}$, and ${}^{(3)}\mathbf{W}^{(c)}$ 
are determined via learning.

	\section{Proposed neural network}
	\subsection{Network structure}
\Figure[!t]()[width=0.95\hsize]{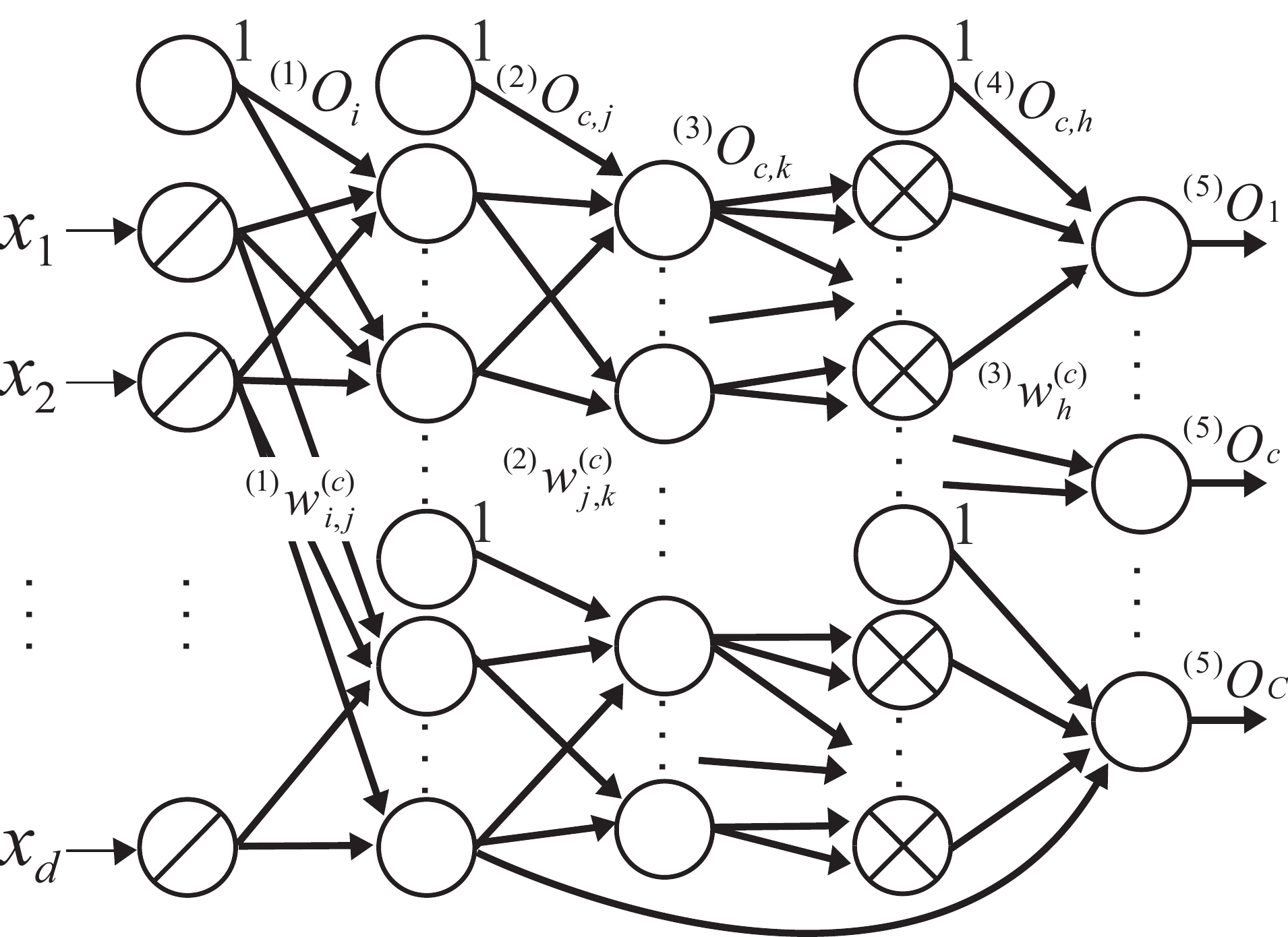}{Structure of the proposed NN. This network is constructed by incorporating the posterior probability calculation based on the Johnson $S_\mathrm{U}$ translation system into the network structure, and consequently consists of five layers. Symbols $\bigcirc$, $\oslash$, and $\otimes$ denote a summation unit, identity unit, and multiplication unit, respectively. The weight coefficients between the first/second layers and the second/third layers correspond to the parameters of the Johnson translation system. The weight coefficients between the fourth/fifth layers correspond to the probabilistic parameters such as the prior probability and the variance matrix. Because of this structure, the output ${}^{(5)}O_{c}$ of this network estimates the posterior probability of each class $c$ given $\bm{x}$, $P(c|\bm{x})$.\label{Structure}}
Fig. \ref{Structure} shows the structure of the proposed NN. This is a five-layer feedforward network
with weight coefficients ${}^{(1)}\mathbf{W}^{(c)}$, ${}^{(2)}\mathbf{W}^{(c)}$, and ${}^{(3)}\mathbf{W}^{(c)}$
between the first/second, second/third, and fourth/fifth layers, respectively.
Symbols $\bigcirc$, $\oslash$, and $\otimes$ denote a summation unit, identity unit, and multiplication unit, respectively. 
Because of this structure, the output ${}^{(5)}O_{c}$ of this network estimates the posterior probability of each class $c$ given $\bm{x}$, $P(c|\bm{x})$.

The first layer consists of $d+1$ units corresponding to the dimensions of the input data $x$.
The relationship between the input and the output is defined as
\begin{IEEEeqnarray}{rCl}
{}^{(1)}I_i &=& \left\{ 
\begin{array}{ll}
1 & (i = 0) \\
x_i & (i = 1,\ldots ,d)
\end{array}
\right., \\
{}^{(1)}O_i &=& {}^{(1)}I_i,
\end{IEEEeqnarray}
where ${}^{(1)}I_i$ and ${}^{(1)}O_i$ are the input and output of the $i$th unit, respectively. 
This layer corresponds to the construction of $\mathbf{X}$ in (\ref{y_trans}).

The second layer is composed of $C(d+1)$ units, each receiving the output of the first layer weighted 
by the coefficient ${}^{(1)}w_{i,j}^{(c)}$.
The relationship between the input ${}^{(2)}I_{c,j}$ and the output ${}^{(2)}O_{c,j}$ of unit 
$\{c,j\}\:(c = 1,\ldots,C$, $j = 1, \ldots, d+1)$ is described as
\begin{IEEEeqnarray}{rCl}
{}^{(2)}I_{c,j} &\!=\!& \sum^{d}_{i=0}{}^{(1)}w_{i,j}^{(c)}{}^{(1)}O_i,\\
{}^{(2)}O_{c,j} &\!=\!& \left\{ 
\begin{array}{ll}
1 \!&\! (j = 0) \\
g({}^{(2)}I_{c,j}) \!&\! (j = 1,\ldots ,d) \\
\sum_{j'=1}^d \log g'({}^{(2)}I_{c,j'}) \!&\! (j = d+1)
\end{array}
\right.\!\!\!\!,
\end{IEEEeqnarray}
where the weight coefficient ${}^{(1)}w_{i,j}^{(c)}$ is an element of the matrix ${}^{(1)}\mathbf{W}^{(c)}$,
which is given as: 
\begin{equation}
{}^{(1)}\mathbf{W}^{(c)} = \left[
\begin{array}{ccc}
{}^{(1)}w_{0,1}^{(c)} & \ldots & {}^{(1)}w_{0,d}^{(c)} \\
{}^{(1)}w_{1,1}^{(c)} &  & \bigzerol \\
 & \ddots &  \\
\bigzerol &  & {}^{(1)}w_{d,d}^{(c)} \\
\end{array} 
\right].
\end{equation}
This layer is equal to the multiplication of ${{}^{(1)}\mathbf{W}^{(c)}}$ and $\mathbf{X}$ in (\ref{y_trans}), 
the construction of $\mathbf{Y}^{(c)}$ in (\ref{z_trans}), and non-coefficient part of Jacobian in (\ref{detJacobain}).

The third layer is comprised of $Cd$ units.
The relationship between the input ${}^{(3)}I_{c,k}$ and the output ${}^{(3)}O_{c,k}$ is defined as
\begin{IEEEeqnarray}{rCl}
{}^{(3)}I_{c,k} &=& \sum^{d}_{j=0}{}^{(2)}w_{j,k}^{(c)}{}^{(2)}O_{c,j},\\
{}^{(3)}O_{c,k} &=& {}^{(3)}I_{c,k},
\end{IEEEeqnarray}
where the weight coefficient ${}^{(2)}w_{j,k}^{(c)}$ is an element of the matrix ${}^{(2)}\mathbf{W}^{(c)}$,
which can be written as: 
\begin{equation}
{}^{(2)}\mathbf{W}^{(c)} = \left[
\begin{array}{ccc}
{}^{(2)}w_{0,1}^{(c)} & \ldots & {}^{(2)}w_{0,d}^{(c)} \\
{}^{(2)}w_{1,1}^{(c)} &  & \bigzerol \\
 & \ddots &  \\
\bigzerol &  & {}^{(2)}w_{d,d}^{(c)} \\
\end{array} 
\right].
\end{equation}
This layer corresponds to the multiplication of ${{}^{(2)}\mathbf{W}^{(c)}}$ and $\mathbf{Y}^{(c)}$ in (\ref{z_trans}).

The fourth layer has $CH$ $(H = 1+ \frac{d(d+1)}{2})$ units.
The relationship between the input ${}^{(4)}I_{c,h}$ and the output ${}^{(4)}O_{c,h}$ of units 
$\{c,h\}$ $(h=1,\ldots,H)$ is defined as
\begin{IEEEeqnarray}{rCl}
\label{input4}
{}^{(4)}I_{c,h}
&=&\left\{ 
\begin{array}{l}
\hspace{-1mm}1 
\hspace{4mm}(h = 1) \\ 
\hspace{-1mm}{}^{(3)}O_{c,k}{}^{(3)}O_{c,k'}\\
\hspace{4.5mm}(h \!=\! k'\!-\!\frac{1}{2}k^2\!+\!(d\!+\!\frac{1}{2})k\!-\!d\!+\!1)
\end{array}
\right.\!\!,\\
{}^{(4)}O_{c,h} &=& {}^{(4)}I_{c,h},
\end{IEEEeqnarray}
where $k \leq k'$ ($k'$ = 1, $\ldots$, $d$), and (\ref{input4}) corresponds to the nonlinear conversion shown in (\ref{nonlinearTrans}).

Finally, the fifth layer consists of $C$ units, and its input ${}^{(5)}I_{c}$ and output ${}^{(5)}O_{c}$ are
\begin{IEEEeqnarray}{rCl}
{}^{(5)}I_{c} &=& \sum^{H}_{h=1}{}^{(3)}w_{h}^{(c)}{}^{(4)}O_{c,h} + {}^{(2)}O_{c,d+1},\\
{}^{(5)}O_{c} &=& \frac{\exp \left({}^{(5)}I_{c} \right)}{\sum^{C}_{c'=1}\exp \left({}^{(5)}I_{c'} \right)}.
\end{IEEEeqnarray}
The output ${}^{(5)}O_{c}$ corresponds to the posterior probability for class $c$, $P(c|\bm{x})$.
Here, the posterior probability $P(c|\bm{x})$ can be calculated if the NN coefficients 
${}^{(1)}\mathbf{W}^{(c)}$, ${}^{(2)}\mathbf{W}^{(c)}$, and ${}^{(3)}\mathbf{W}^{(c)}$ have been appropriately established.

\subsection{Learning algorithm}
This subsection describes a learning algorithm that can acquire a unique optimal solution without any hyperparameters.
The learning algorithm consists of two steps.

In the first step, we estimate ${}^{(1)}\mathbf{W}^{(c)}$ and ${}^{(2)}\mathbf{W}^{(c)}$, which contain 
the parameters of the Johnson translation system. 
Although various parameter estimation algorithms have been proposed for the Johnson translation system, 
this paper adopts the percentile method \cite{slifker1980johnson}, because it analytically estimates the parameters 
with a certain degree of accuracy. 
The percentile method calculates Johnson system parameters by comparing distances in the tails with distances in the central portion of the distribution. 
Using the percentile method, the parameters for the Johnson translation system are determined as follows: 
\begin{IEEEeqnarray}{rCl}
\delta_i^{(c)} &=& \frac{2z}{\mathrm{cosh}^{-1}\left[\frac{1}{2}\left(\frac{m_i^{(c)}}{p_i^{(c)}}+\frac{n_i^{(c)}}{p_i^{(c)}}\right)\right]}, \\
\gamma_i^{(c)} &=& \delta_i^{(c)}\mathrm{sinh}^{-1}\left[\frac{\frac{n_i^{(c)}}{p_i^{(c)}}-\frac{m_i^{(c)}}{p_i^{(c)}}}{2\left(\frac{m_i^{(c)}n_i^{(c)}}{(p_i^{(c)})^2}-1\right)^{\frac{1}{2}}}\right], \\
\lambda_i^{(c)} &=& \frac{2p_i^{(c)}\left(\frac{m_i^{(c)}n_i^{(c)}}{(p_i^{(c)})^2}-1\right)^{\frac{1}{2}}}
{\left(\frac{m_i^{(c)}}{p_i^{(c)}}+\frac{n_i^{(c)}}{p_i^{(c)}}-2\right)\left(\frac{m_i^{(c)}}{p_i^{(c)}}+\frac{n_i^{(c)}}{p_i^{(c)}}+2\right)^{\frac{1}{2}}}, \\
\xi_i^{(c)} &=& \frac{x_{z,i}^{(c)} + x_{-z,i}^{(c)}}{2} + \frac{p_i^{(c)}\left(\frac{n_i^{(c)}}{p_i^{(c)}}-\frac{m_i^{(c)}}{p_i^{(c)}}\right)}{2\left(\frac{m_i^{(c)}}{p_i^{(c)}}+\frac{n_i^{(c)}}{p_i^{(c)}}-2\right)},
\end{IEEEeqnarray}
where $z > 0$ is chosen depending on the number of data points, 
and $i = 1, \ldots, d$, $c = 1, \ldots, C$ are indices corresponding to the dimension and class, respectively.
The variable $x_{\zeta,i}^{(c)}$ ($\zeta = -3z, -z, z, 3z$) is the $P_\zeta$th percentile of the $i$th dimension of training data for class $c$, 
where $P_\zeta$ is the percentage of the area in the normal distribution corresponding to $\zeta$.
Using percentiles, $m_i^{(c)}, n_i^{(c)}, p_i^{(c)}$ are calculated as: 
\begin{IEEEeqnarray}{rCl}
m_i^{(c)} &=& x_{3z,i}^{(c)} - x_{z,i}^{(c)}, \\
n_i^{(c)} &=& x_{-z,i}^{(c)} - x_{-3z,i}^{(c)}, \\
p_i^{(c)} &=& x_{z,i}^{(c)} - x_{-z,i}^{(c)}.
\end{IEEEeqnarray}
For more detail, refer to \cite{slifker1980johnson}.
${}^{(1)}\mathbf{W}^{(c)}$ and ${}^{(2)}\mathbf{W}^{(c)}$ can then be determined 
by substituting $\delta_i^{(c)}, \gamma_i^{(c)}, \lambda_i^{(c)}$, and $\xi_i^{(c)}$ in (\ref{y_trans}) and (\ref{z_trans}).

The second step concerns the discriminative learning of the remaining weight ${}^{(3)}\mathbf{W}^{(c)}$, 
which includes probabilistic parameters such as the prior probability $P(c)$ and covariance matrix ${\bf{\Sigma}}^{(c)}$.
A set of vectors $\bm{x}^{(n)}$ $(n = 1, \ldots, N)$ is given for training, with the teacher vector
$\mathbf{T}^{(n)} = [T^{(n)}_1, \ldots, T^{(n)}_c,$ $\ldots, T^{(n)}_C]$ for the $n$th input.
The training process of ${}^{(3)}\mathbf{W}^{(c)}$ involves minimizing the energy function $E$, which is defined as
\begin{equation}
E = \sum^{N}_{n=1}E_n = -\sum^{N}_{n=1}\sum^{C}_{c=1}T^{(n)}_c\log {}^{(5)}O_{c}^{(n)},
\end{equation}
to maximize the log-likelihood.
Here, ${}^{(5)}O_{c}^{(n)}$ is the output for an input vector $\bm{x}^{(n)}$.
The weight modification for ${}^{(3)}w_{h}^{(c)}$
based on Newton's method is defined as
\begin{IEEEeqnarray}{rCl}
\label{w3}
{}^{(3)}\mathbf{W}_\mathrm{new} &=& {}^{(3)}\mathbf{W}_\mathrm{old} -\mathbf{H}^{-1}\nabla E,
\end{IEEEeqnarray}
where ${}^{(3)}\mathbf{W}_\mathrm{old}$ and ${}^{(3)}\mathbf{W}_\mathrm{new}$ are the weight coefficients before and after weight modification, 
which have ${}^{(3)}\mathbf{W}^{(c)}$ in the $c$th block. 
$\nabla E_n$ is the gradient vector whose $h$th element in the $c$th block can be calculated as 
\begin{IEEEeqnarray}{rCl}
\frac{\partial E}{\partial {}^{(3)}w_{h}^{(c)}} 
&=& \sum^{N}_{n=1}\frac{\partial E_n}{\partial {}^{(3)}w_{h}^{(c)}} \nonumber \\
&=& \sum^{N}_{n=1}\sum^{C}_{c'=1}\!\frac{\partial E_n}{\partial {}^{(5)}O_{c'}^{(n)}}
					  \frac{\partial {}^{(5)}O_{c'}^{(n)}}{\partial {}^{(5)}I_{c}^{(n)}}
					  \frac{\partial {}^{(5)}I_{c}^{(n)}}{\partial {}^{(3)}w_{h}^{(c)}}\nonumber \\
&=& \sum^{N}_{n=1}({}^{(5)}O_{c}^{(n)} - T^{(n)}_{c}){}^{(4)}O_{c,h}^{(n)}.
\end{IEEEeqnarray}
$\bf H$ is the Hessian matrix comprised of $H \times H$ blocks, 
where the $(h, l)$ element of block $(c, k)$ is 
\begin{IEEEeqnarray}{rCl}
\label{HessianComponent}
\lefteqn{\sum^{N}_{n=1}\frac{\partial ^2 E_n}{\partial {}^{(3)}w_{h}^{(c)}\partial{}^{(3)}w_{l}^{(k)}}} \nonumber \\
&=& \sum^{N}_{n=1}{}^{(5)}O_{k}^{(n)}(\delta_{c, k} - {}^{(5)}O_{c}^{(n)}){}^{(4)}O_{c,h}^{(n)}{}^{(4)}O_{k,l}^{(n)}.
\end{IEEEeqnarray}
Note that $\bf H$ is positive semi-definite (see Appendix A). 
It follows that $E$ is a convex function of ${}^{(3)}\mathbf{W}^{(c)}$, 
and hence has a unique minimum.

Using this algorithm, the process of training the network converges to a unique solution
without the need for any hyperparameters.

	\section{Simulation experiment}
	\subsection{Method}
To verify that the proposed network can properly calculate the posterior probability for data with skewness and kurtosis, 
we performed a simulation experiment using two-dimensional ($d=2$) two-class ($C=2$) data. 
The data were artificially generated using the inverse of the multivariate Johnson $S_\mathrm{U}$ translation system \cite{stanfield1996multivariate}. 
Table \ref{parameters} lists the parameters used for each class in this generation.
\begin{table}[t]
	\caption{Parameters for data generation in the simulation experiment}
	\label{parameters}
	\centering
	\begin{tabular}{llllll} \hline
							 & $\xi$ 	& $\lambda$ & $\delta$ 	& $\gamma$ 	& $\Sigma^{-1}$\\	\hline
	\multirow{2}{*}{Class 1} & 0.15		& 0.04 		& 0.9 		& -0.9 		& \multirow{2}{*}{0.6}	\\
							 & 0.7 		& 0.05 		& 0.8 		& 0.5 		& \\	\hline
	\multirow{2}{*}{Class 2} & 0.5 		& 0.05 		& 0.8 		& 0.5 		& \multirow{2}{*}{0.9}	\\
							 & 0.55 	& 0.01 		& 0.5 		& -0.5 		& \\	\hline
	\end{tabular}
\end{table}
An example of a dataset used in the experiments is shown in Fig. \ref{ArtificialData}.

\Figure[!t]()[width=0.95\hsize]{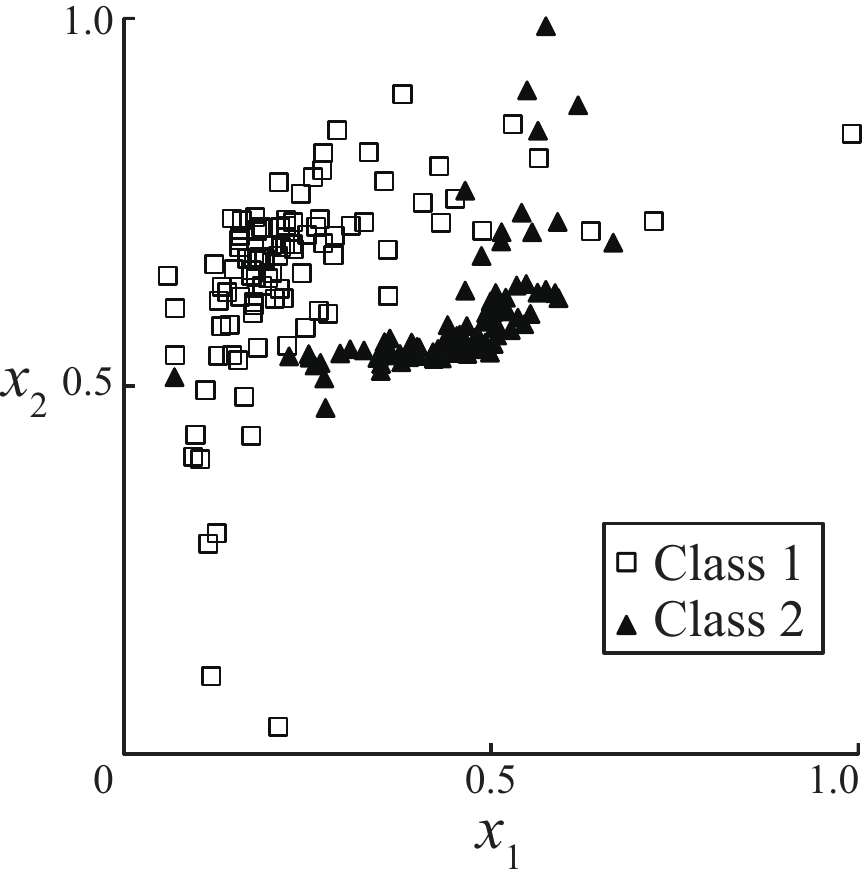}
   {Scattergram of a dataset used in the simulation experiment. Each class has different skewness and kurtosis, and they should be considered to accurately calculate the posterior probabilities.\label{ArtificialData}}
Each class has different skewness and kurtosis qualities, as well as a different mean and variance.

In the experiment, 100 samples were treated as training samples for each class. 
The function $g_i(y)$ was of type $S_\mathrm{U}$ (unbounded). 
After training, the proposed NN was tested using inputs in the range $0 \leq x_1 \leq 1$ and $0 \leq x_2 \leq 1$. 
The corresponding posterior probabilities were compared with those given by LLR and
LLGMN \cite{tsuji1999log}. 
The LLR was trained using Newton's method \cite{bishop2006pattern}, 
and LLGMN was trained by terminal learning \cite{zak1989terminal} with an ideal convergence time of 1.0 and a learning sampling time of 0.001. 
The number of components $M_c$ in the LLGMN was varied from 1 to 10.


\Figure[!t]()[width=0.95\hsize]{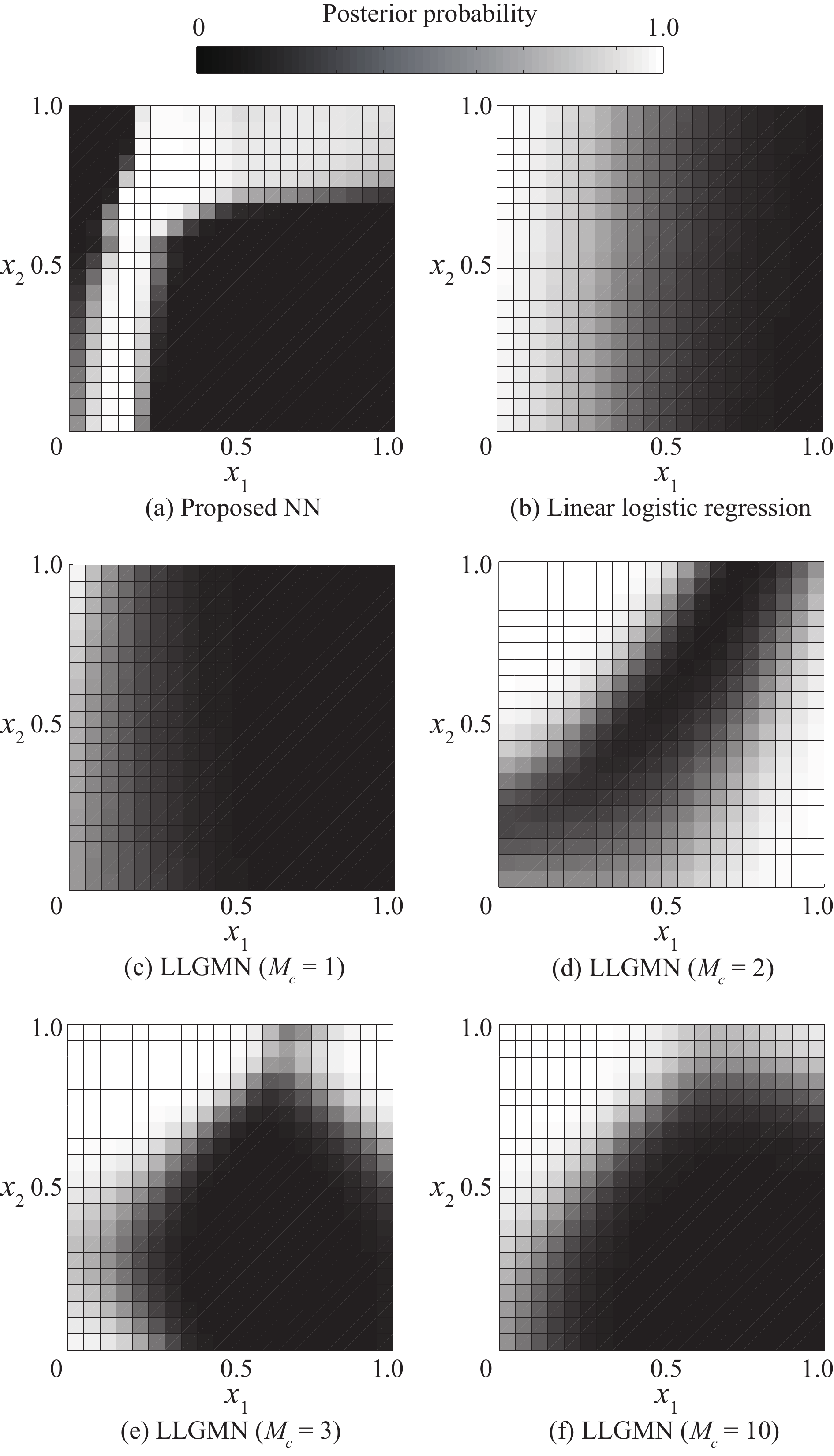}{Posterior probability of class 1 ($P(c=1|\bm{x})$) given by (a) the proposed NN, (b) LLR, and (c) LLGMN ($M_c=1$), (d) LLGMN ($M_c=2$), (e) LLGMN ($M_c=3$), and (f) LLGMN ($M_c=10$), where $M_c$ is the number of components for a Gaussian mixture model used in the LLGMN. The probability of class 2 ($P(c=2|\bm{x})$) is the reverse of that of class 1 with respect to black and white.\label{PPD}}

\subsection{Results and discussion}
Fig. \ref{PPD} shows the posterior probability of class 1 ($P(c=1|\bm{x})$) given by the proposed NN, LLR, and LLGMN. 
The probability of class 2 ($P(c=2|\bm{x})$) is clear from this graph, because it can be calculated as $P(c=2|\bm{x}) = 1 - P(c=1|\bm{x})$.

From Fig. \ref{PPD} (a), it is clear that the posterior probability given by the proposed NN resembles the distribution shape of the experimental data of class 1 (Fig. \ref{ArtificialData}). 
The probability given by LLR (Fig. \ref{PPD} (b)) is totally different from the experimental data distribution. 
Although LLGMN with $M_c=1$ (Fig. \ref{PPD} (c)) is also different from the experimental data distribution, 
this classifier produces probability that becomes closer to the experimental data as the number of components increases. 

It can be inferred that the proposed NN is capable of appropriately dealing with data including skewness and kurtosis, 
because it is based on the Johnson translation system.
In contrast, LLR cannot be adapted to data with skewness and kurtosis, because it is a linear classifier.
LLGMN is capable of handling data with skewness and kurtosis 
when the number of components is sufficiently large, 
but does not represent the data distribution well with few components. 
The above results demonstrate that the proposed NN can handle data including skewness and kurtosis without hyperparameters, 
although conventional methods require a hyperparameter optimization step.

	\section{EMG classification experiment}
	\subsection{Method}
To evaluate the suitability of the proposed network for real biological data, 
a classification experiment was conducted using EMG data. 
Details of the data acquisition are described in the next subsection.
Table \ref{DataDescription} shows the characteristics of six datasets prepared for this experiment 
in terms of the number of motions that the subjects performed, number of electrodes, 
number of subjects, number of trials for each subject, and number of samples for each trial.
\begin{table}[t]
	\caption{Summary of dataset for the EMG classification experiment}
	\begin{tabular}{llllll} \hline
	Dataset & \# Motions & \# Electrodes 	& \# Sub. 	& \# Trials	 	& \# Samples\\ \hline
	I 		& 6 		 & 6 		    	& 1 		& 10			& 20000\\
	II 	    & 14 	     & 8 		    	& 8 		& 4			    & 36000\\
	III 	& 15 		 & 8 		    	& 8 		& 3			    & 72000\\
	IV 	    & 16 		 & 13 		        & 1 		& 15			& 10000\\
	V 		& 17 		 & 12 		        & 9	    	& 6		    	& 12000\\
	VI	    & 6 		 & 2 			    & 5		    & 30			& 2000\\ \hline
	\end{tabular}
	\label{DataDescription}
\end{table}
The number of motions and the number of electrodes correspond to 
the number of classes and the number of input dimensions, respectively. 
The training samples were randomly chosen from the available samples for each trial, 
with the remaining samples used for testing. 
Because it is difficult to procure many training samples in real-world applications, 
only 1\% of the available samples were selected for training to evaluate the validity of the proposed NN for learning with limited training data. 

We compared the performance of the proposed NN with that of {$\nu$}-SVM \cite{chang2001training} with a one-vs-one classifier, 
LLGMN \cite{tsuji1999log}, MLP, 
LLR, and $k$-NN. 
The hyperparameters of {$\nu$}-SVM ($\gamma$ and $\nu$) were optimized by 10-fold cross-validation (CV) and a 10 $\times$ 10 grid search 
($\gamma$ ranging from $\log_{10}5.0$ to $\log_{10}1.0^{-5}$, and $\nu$ ranging from $\nu_\mathrm{max}$ to $\log_{10}1.0^{-5}$ at even intervals in logarithmic space, 
where $\nu_\mathrm{max}$ is dependent on the ratio of labels in the training data). 
LLGMN was trained by terminal learning \cite{zak1989terminal} with an ideal convergence time of 1.0 and a learning sampling time of 0.001. 
The number of components (from 1 to 5) in the LLGMN was determined using 10-fold CV. 
The number of nodes (from $d$ to $d+10$) in the hidden layer of MLP was also determined using 10-fold CV, 
and MLP was trained using the back propagation algorithm with a learning rate of 0.1.
The LLR was trained using Newton's method, 
and the value of $k$ in the $k$-NN algorithm was chosen in the range 1 to 10 using 10-fold CV. 
All algorithms were programmed using C++ and the dlib C++ Library \cite{king2009dlib}. 
The experiments were run on a computer with an Intel Core(TM) i7-3770K (3.5 GHz) processor and 16.0 GB RAM for Datasets I--V, 
and an Intel Core(TM) i7-7700K (4.2 GHz) processor and 16.0 GB RAM for Datasets VI.

To evaluate the usefulness of a classifier for real-world applications, 
it is necessary to measure not only the classification accuracy, but also the training/preparation time 
and the prediction time. 
We therefore compared the performance of the above algorithms through four metrics: 
accuracy, CV time, training time, and prediction time. 
Accuracy is defined as $100 \times N_\mathrm{correct} / N_\mathrm{total}$, where $N_\mathrm{correct}$ is the number of correctly classified test samples 
and $N_\mathrm{total}$ is the total number of test samples. 
CV time is the total time taken for hyperparameter optimization based on CV,
and training time is the time required for training. 
The sum of CV time and training time can be considered as the time until the classifier becomes available. 
Prediction time is the total time taken to classify all test samples.
These metrics were measured for each trial and each subject, and the average value was then calculated.

\subsection{Data acquisition}
Dataset I contains six-channel ($d=6$) EMG data recorded by the authors. 
The six pairs of electrodes were located as follows: 
Ch. 1: extensor carpi ulnaris; Ch. 2: flexor digitorum profundus; 
Ch. 3: extensor digitorum; Ch. 4: flexor carpi ulnaris; 
Ch. 5: triceps brachii; Ch. 6: biceps brachii (see Fig. \ref{chLocation}).

\Figure[!t]()[width=0.95\hsize]{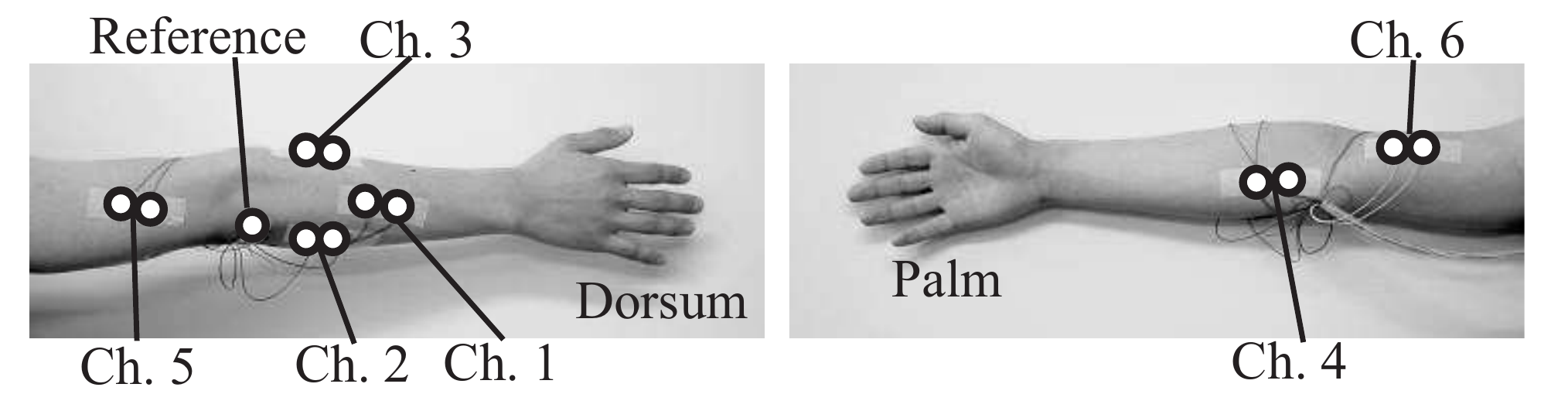}{Locations of electrodes for dataset I\label{chLocation}}

The healthy 22-year-old male subject performed six successive motions in a relaxed state 
($C = 6$; M1: hand opening; M2: hand grasping; M3: wrist extension; M4: wrist flexion; M5: pronation; M6: supination).
EMG signals were recorded at 1 kHz and digitalized using a 16-bit A/D converter. 
Feature extraction was then conducted according to the method of \cite{fukuda2003human}.
The signals were rectified and smoothed using a second-order Butterworth low-pass filter with a cut-off frequency of 1 Hz. 
These features were defined as $EMG_i(n)$ $(i = 1, \ldots , d, n= 1, \ldots, N$: $N$ is the number of data) and normalized as follows: 
\begin{equation}
\label{norm}
	x_i^{(n)} = \frac{EMG_i(n)-\overline{EMG_i^\mathrm{st}}}{\displaystyle\sum^{d}_{i'=1}\left(EMG_{i'}(n)-\overline{EMG_{i'}^\mathrm{st}}\right)},
\end{equation}
where $\overline{EMG_i^\mathrm{st}}$ is the mean of $EMG_i(n)$ 
in a state of muscular relaxation. 
$x_i^{(n)}$ was then used as the input for the network. 

Datasets II and III are those used in \cite{khushaba2013muscle} and \cite{khushaba2012electromyogram}, respectively\footnote{These datasets are available at Dr. Khushaba's webpage: http://www.rami-khushaba.com/electromyogram-emg-repository.html}. 
Dataset II contains measurements from eight subjects (aged from 20--35 years old) while seated on an armchair, putting their hands on a steering wheel attached to a desk 
and performing twelve classes of finger pressures and two classes of finger pointing (i.e., a total of 14 classes ($C = 14$)).
For dataset III, eight subjects (aged from 20--35 years old) performed fifteen classes of finger and hand movements 
while seated on an armchair with their arm supported and fixed in one position.
In both these datasets, EMG signals were recorded using eight-channel electrodes at 4 kHz and digitalized using a 12-bit A/D converter.
Feature extraction for these datasets was conducted by rectification and smoothing, as for dataset I.

Dataset IV contains measurements
from a healthy 23-year-old male performing sixteen forearm motions ($C = 16$). 
EMG signals were recorded using thirteen pairs of electrodes ($d = 13$) at 1 kHz with a 60-Hz notch filter and a bandpass filter of 0.1--200 Hz.
Details of the experimental conditions are described in \cite{shibanoki2013quasi}.
Additionally, to evaluate the performance of the proposed NN for more difficult classification problems, 
we also confirmed the change of accuracy according to the decrease of the number of electrodes.
The elimination of electrodes was conducted in order from the ones having the largest channel numbers, 
and then average classification accuracy and standard deviation were calculated for all the trials. 

Dataset V is the Ninapro Database 3 exercise 1 \cite{atzori2014electromyography}, which is available on Ninaweb\footnote{http://ninapro.hevs.ch/node/131}. 
EMG signals were recorded from 11 trans-radial amputated subjects using 12 electrodes ($d=12$) at 2 kHz while conducting 17 movements ($C=17$).
Each movement lasted five seconds and was repeated six times with a rest interval of three seconds. 
Because some electrodes were missed for two subjects, 
nine out of the 11 subjects were used to uniform the number of channels
in the classification experiment.
Feature extraction for these datasets was conducted by rectification and smoothing, as for other datasets.

Dataset VI is the sEMG for Basic Hand movements Data Set provided by Sapsanis {\it et al.} \cite{sapsanis2013improving}. 
Five healthy subjects (two males and three females) were asked to perform six grasping movements ($C=6$): 
holding a cylindrical tool, 
supporting a heavy load, 
holding a small tool, 
grasping with palm facing the object, 
holding a spherical tool, 
and holding a thin and flat object.
Each movement lasted six seconds and was repeated 30 times. 
EMG signals were collected from two forearm surface EMG electrodes ($d=2$) at a sampling rate of 500 Hz. 
Feature extraction was conducted in the same way as for other datasets, 
and the initial 1,000 samples for each movement were then discarded to remove transition states. 
For this dataset, classification accuracy is calculated based on $5 \times 2$ CV approach referring to the original paper that provides this dataset \cite{sapsanis2013improving}. 
The number of training data is limited also in this dataset by randomly sampling 1\% of the training set in each fold.

\subsection{Results}
Table \ref{ResultEMG} summarizes the results of EMG classification. 
Values are the average and standard deviation of scores measured for each trial of each subject, 
and are presented as ``average value $\pm$ standard deviation'' or ``average value'' if the standard deviation was 0.
``$\ast\ast$'' in the accuracy column denotes a significant difference, based on the Holm method, between that algorithm and the proposed NN ($p < 0.01$). 
The absence of ``$\ast\ast$'' in accuracy denotes no significant statistical difference. 
\begin{table*}[t]
	\caption{Results of EMG classification}
	\centering
	\begin{threeparttable}
	\begin{tabular}{llllll}\hline
	Dataset & Algorithm 	& Accuracy [\%]			& CV time [s]				& Training time [s] 	& Prediction time [s]	\\ \hline
    I       & JohnsonNN     & $100$                 & $0$                       & $0.370 \pm 0.015$     & $0.525 \pm 0.004$\\
            & SVM           & $100$                 & $23.390 \pm 0.548$        & $0.115 \pm 0.017$     & $29.221 \pm 7.102$\\
            & LLGMN         & $100$                 & $197.497 \pm 1.187$       & $1.209 \pm 0.011$     & $0.318 \pm 0.022$\\
            & MLP           & $100$                 & $207.315 \pm 7.259$       & $36.034 \pm 0.58$     & $0.07 \pm 0.004$\\
            & LLR           & $99.504 \pm 1.489$    & $0$                       & $0.029 \pm 0.007$     & $0.303 \pm 0.037$\\
            & $k$-NN        & $100$                 & $1.347 \pm 0.037$         & $0$                   & $15.231 \pm 0.149$\\
            & RandomForest  & $100$                 & $4.084 \pm 0.079$         & $0.206 \pm 0.006$     & $2.562 \pm 0.146$\\ \hline
    II      & JohnsonNN     & $100$                 & $0$                       & $29.025 \pm 4.891$    & $6.536 \pm 0.104$\\
            & SVM           & $99.999 \pm 0.005$    & $522.975 \pm 8.864$       & $0.447 \pm 0.349$     & $131.309 \pm 119.874$\\
            & LLGMN         & $99.198 \pm 0.695$ ** & $3074.678 \pm 137.516$    & $40.3 \pm 24.863$     & $4.392 \pm 1.166$\\
            & MLP           & $96.689 \pm 5.300$ ** & $1609.333 \pm 34.906$     & $161.952 \pm 3.923$   & $0.418 \pm 0.034$\\
            & LLR           & $95.908 \pm 6.812$ ** & $0$                       & $1.145 \pm 0.619$     & $2.792 \pm 0.131$\\
            & $k$-NN        & $99.999 \pm 0.002$    & $26.992 \pm 0.779$        & $0$                   & $319.459 \pm 11.172$\\
            & RandomForest  & $98.862 \pm 1.643$ ** & $43.713 \pm 0.674$        & $1.379 \pm 0.024$     & $19.971 \pm 1.645$\\ \hline
    III     & JohnsonNN     & $99.973 \pm 0.107$    & $0$                       & $99.295 \pm 39.217$   & $15.235 \pm 0.324$\\
            & SVM           & $99.726 \pm 0.633$    & $2454.33 \pm 49.024$      & $0.922 \pm 0.614$     & $254.447 \pm 183.636$\\
            & LLGMN         & $96.131 \pm 3.286$ ** & $6844.277 \pm 168.153$    & $105.441 \pm 55.835$  & $10.436 \pm 2.686$\\
            & MLP           & $83.631 \pm 8.659$ ** & $3483.782 \pm 47.591$     & $356.209 \pm 7.832$   & $1.021 \pm 0.119$\\
            & LLR           & $90.590 \pm 8.986$ ** & $0$                       & $2.329 \pm 1.951$     & $6.904 \pm 0.382$\\
            & $k$-NN        & $99.997 \pm 0.005$    & $144.351 \pm 6.757$       & $0$                   & $1740.607 \pm 62.525$\\
            & RandomForest  & $97.252 \pm 3.783$    & $105.956 \pm 2.710$       & $3.357 \pm 0.074$     & $48.156 \pm 3.105$\\ \hline
    IV      & JohnsonNN     & $100$                 & $0$                       & $66.226 \pm 6.118$    & $4.149 \pm 0.037$\\
            & SVM           & $100$                 & $70.399 \pm 1.417$        & $0.144 \pm 0.071$     & $66.870 \pm 66.167$\\
            & LLGMN         & $98.386 \pm 2.022$ ** & $2638.634 \pm 48.720$     & $26.279 \pm 18.556$   & $2.080 \pm 0.850$\\
            & MLP           & $85.038 \pm 11.362$ **& $786.566 \pm 9.134$       & $55.727 \pm 1.386$    & $0.181 \pm 0.014$\\
            & LLR           & $97.906 \pm 1.863$ ** & $0$                       & $0.542 \pm 0.527$     & $1.042 \pm 0.096$\\
            & $k$-NN        & $100$                 & $3.214 \pm 0.087$         & $0$                   & $36.481 \pm 0.357$\\
            & RandomForest  & $100$                 & $31.526 \pm 0.394$        & $0.389 \pm 0.017$     & $5.136 \pm 0.168$\\ \hline
    V       & JohnsonNN     & $39.122 \pm 7.568$    & $0$                       & $218.288 \pm 19.043$  & $6.511 \pm 0.056$\\
            & SVM           & $39.169 \pm 7.699$    & $124.734 \pm 1.403$       & $0.065 \pm 0.014$     & $23.799 \pm 5.279$\\
            & LLGMN         & $34.782 \pm 7.297$    & $2965.718 \pm 76.741$     & $40.293 \pm 19.812$   & $3.025 \pm 0.937$\\
            & MLP           & $36.622 \pm 6.492$    & $960.015 \pm 6.262$       & $73.066 \pm 1.389$    & $0.274 \pm 0.030$\\
            & LLR           & $33.848 \pm 8.373$    & $0$                       & $0.369 \pm 0.098$     & $1.291 \pm 0.044$\\
            & $k$-NN        & $37.798 \pm 7.455$    & $4.974 \pm 0.295$         & $0$                   & $58.331 \pm 1.936$\\
            & RandomForest  & $33.109 \pm 8.264$    & $39.456 \pm 0.677$        & $0.569 \pm 0.032$     & $8.446 \pm 0.767$\\ \hline
    VI      & JohnsonNN     & $68.190 \pm 10.014$   & $0$                       & $0.049 \pm 0.032$     & $0.501 \pm 0.017$\\
            & SVM           & $67.544 \pm 9.073$    & $31.037 \pm 0.603$        & $0.064 \pm 0.041$     & $7.821 \pm 5.088$\\
            & LLGMN         & $65.852 \pm 6.457$    & $92.514 \pm 2.071$        & $1.609 \pm 0.637$     & $0.505 \pm 0.066$\\
            & MLP           & $53.117 \pm 7.198$ ** & $171.2 \pm 2.574$         & $56.152 \pm 0.988$    & $0.115 \pm 0.013$\\
            & LLR           & $70.992 \pm 8.646$    & $0$                       & $0.02 \pm 0.006$      & $0.454 \pm 0.025$\\
            & $k$-NN        & $68.039 \pm 8.658$    & $2.766 \pm 0.115$         & $0$                   & $29.957 \pm 0.669$\\
            & RandomForest  & $69.388 \pm 9.645$    & $1.321 \pm 0.073$         & $0.445 \pm 0.042$     & $5.349 \pm 0.568$\\ \hline
	\end{tabular}
	\begin{tablenotes}
	\item **: significant difference with the proposed NN ($p < 0.01$)
	\end{tablenotes}
	\end{threeparttable}
	\label{ResultEMG}
\end{table*}
Fig. \ref{ConfusionMatrix} shows the confusion matrix of the classification results for the dataset VI using the proposed NN. 

\Figure[!t]()[width=0.95\hsize]{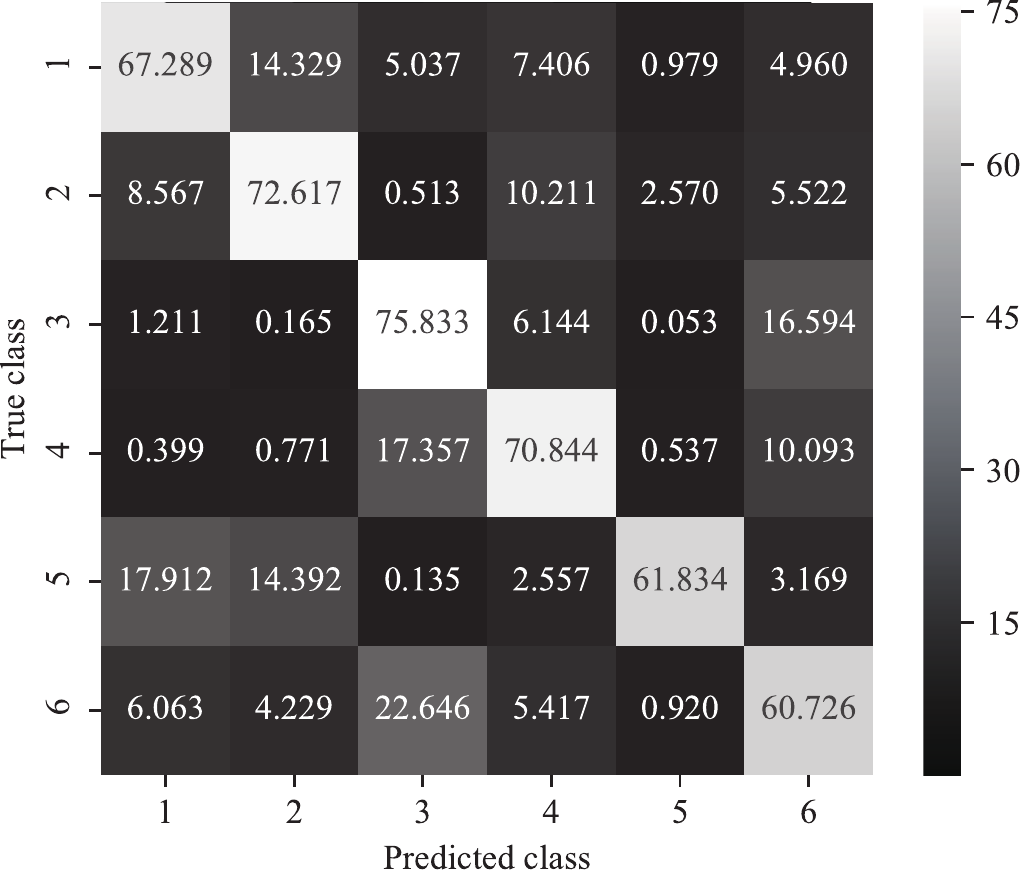}{Confusion matrix of the classification results for the dataset VI. Values are normalized by the number of test samples for each class.\label{ConfusionMatrix}}

Fig. \ref{DimReduction} shows accuracy for each number of electrodes, while decreasing the electrodes for dataset IV. 
For comparison, the accuracies of $\nu$-SVM are also plotted. 
Significant differences between the proposed NN and $\nu$-SVM were confirmed when the number of electrodes was $d=2$ and $d=3$ ($p < 0.05$).

\Figure[!t]()[width=0.95\hsize]{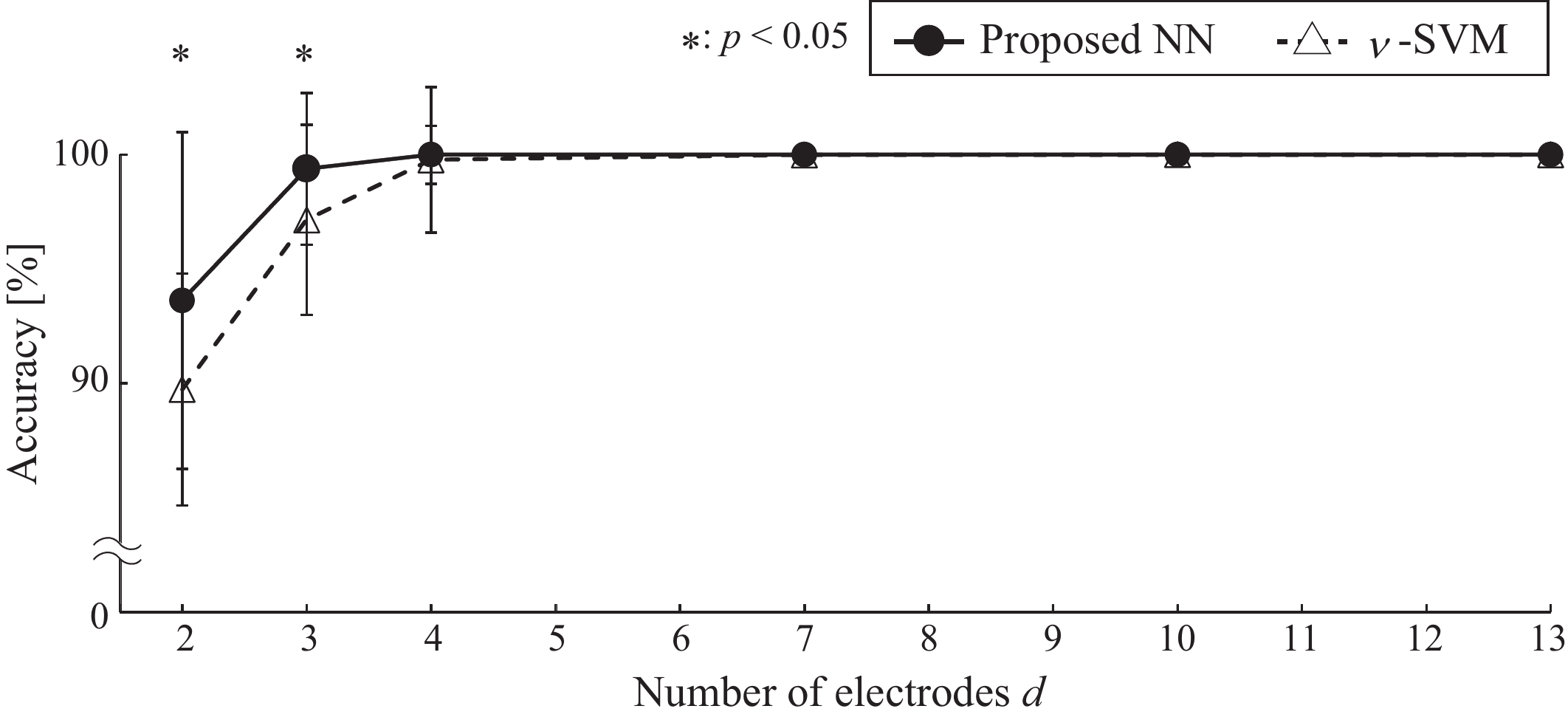}{Classification accuracy for each number of electrodes (Dataset IV).\label{DimReduction}}


Fig. \ref{AccVsTime} shows the relationship between accuracy and preparation time (CV time + training time), 
representing the time until the classifier becomes available, 
and between accuracy and prediction time for each classification method. Proximity to the upper-left corner indicates superior performance. 

\Figure[!t]()[width=0.95\hsize]{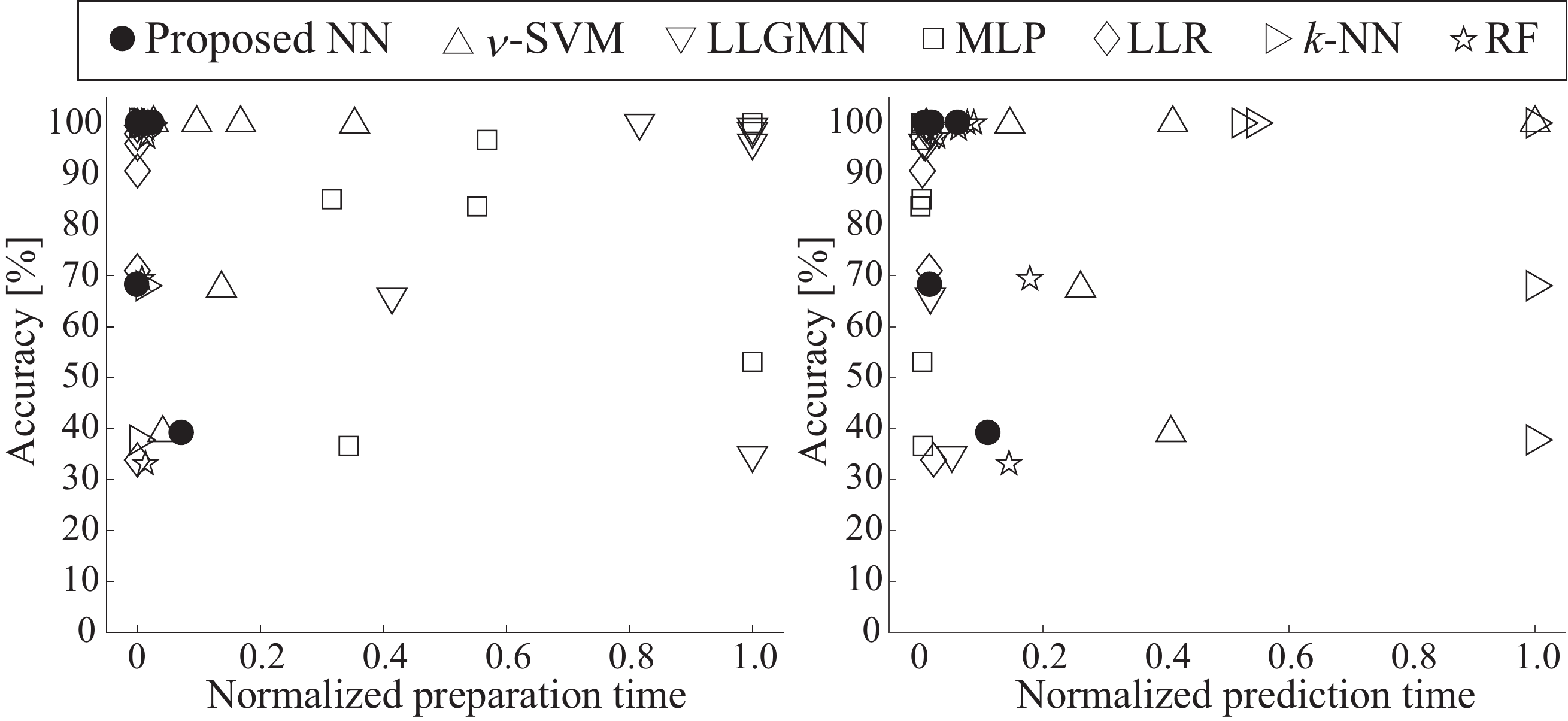}{Relationships between (left) accuracy and preparation time (CV time + Training time) and between (right) accuracy and prediction time by using each classification method for six different datasets. The preparation time and the prediction time are normalized by the maximum value for each dataset. Proximity to the upper-left corner indicates superior performance.\label{AccVsTime}}

\subsection{Discussion}
In terms of accuracy, the proposed NN, {$\nu$}-SVM, and $k$-NN achieve the same level of performance, 
demonstrating a suitability for EMG signal classification. 
The proposed NN involves Johnson distribution in its structure based on prior knowledge of EMG signals 
for appropriate modeling of EMG distribution in the network.
{$\nu$}-SVM showed strong generalization ability derived from margin maximization. 
$k$-NN can be used to express arbitrary complex decision boundaries based on determination of the parameter $k$, 
ensuring a fit to the skewness and kurtosis of EMG data.
On the other hand, the accuracies of LLGMN, MLP, and LLR were notably low in some cases. 
This is because LLGMN and MLP require many parameters to fit data with skewness and kurtosis, 
which results in over-fitting, 
and LLR cannot solve nonlinear classification problems because it is a linear classifier. 
For dataset V, accuracies were relatively low in all of the algorithms. 
This is because this dataset is recorded from amputated subjects, 
and therefore EMG signals were unstable and the reproducibility of motions was low 
compared with the data recorded form intact subjects.

In Fig. \ref{ConfusionMatrix}, confusion between class 3 and class 6  is relatively frequent compared with other classes. 
This is reasonable since the motions of these classes are similar (class 3: holding a small tool, class 6: holding a thin and flat object). 
However, there was no extreme bias toward a certain class; therefore the proposed NN worked properly for multi-class classification.

In Fig. \ref{DimReduction}, the accuracies of the proposed NN and $\nu$-SVM both decreased according to the decrease of the number of electrodes.
This is because the decrease of electrodes yielded the loss of information enough to classify the motions, 
making the classification problem difficult. 
In particular, the accuracies were sharply reduced when the number of channels was reduced $d=3$ to $d=2$, although the proposed NN exceeded $\nu$-SVM. 
One possible explanation is that the substantial reduction of the input dimensions yielded the overlap of distribution for each class, 
and thus the proposed NN could not model the data distribution precisely.

With respect to CV time, LLGMN and MLP took particularly long. 
In contrast, the proposed NN and LLR had CV times of 0
because they have a unique solution of learning and therefore do not require hyperparameters such as a learning rate. 

The training time for the proposed NN was relatively short for dataset I. 
In datasets II, III, IV, and V, however, significant training time was required. 
This is because the cost of calculating the Hessian matrix and finding its inverse (see (\ref{w3}) and (\ref{HessianComponent})) increases 
with the number of classes and input dimensions. 
Although the overall time until the classifier becomes available is relatively short (because the CV time is 0), 
there is room for improvement by making the numerical calculations more efficient.

Regarding the prediction time, {$\nu$}-SVM and $k$-NN took particularly long.
This is because {$\nu$}-SVM was originally a binary classifier, and thus solves multi-class classification problems 
by calculating all combinations of two-class classification. 
$k$-NN also has a long computation time, because it calculates the distance between the input sample and every training sample. 
The prediction time of the proposed NN is relatively short because it realizes a compact model for EMG classification 
by incorporating prior knowledge of the processed EMG characteristics.

Overall performance is summarized in Fig. \ref{AccVsTime}. 
The plots of the proposed NN are concentrated toward the upper-left corner, 
demonstrating well-balanced performance for accuracy and computation cost. 

Finally, the performance of the proposed NN can be summarized as follows: 
\begin{itemize} 
\item High accuracy for classification of EMG signals
\item Relatively short time until the classifier becomes available 
\item Shorter prediction time than {$\nu$}-SVM and $k$-NN
\end{itemize}

	\section{Conclusion}
	In this paper, we proposed a NN based on the Johnson $S_\mathrm{U}$ translation system. 
The NN includes a discriminative model based on the multivariate Johnson $S_\mathrm{U}$ translation system, 
with the model transformed into linear combinations of weight coefficients and nonlinearly transformed input vectors. 
This enables the representation of more flexible distributions for data with skewness and kurtosis. 
Parameters describing the shape of the distribution can be determined as network coefficients via network learning. 
The proposed NN can be trained without hyperparameter optimization, and the training converges to a unique solution. 
In addition, the posterior probability of input vectors for each class can be calculated as the output of the NN.

In a simulation experiment, the proposed network was shown to be more suitable than a conventional GMM-based network and linear logistic regression
for data with skewness and kurtosis.
The applicability of the proposed NN to biosignal classification was also demonstrated by the results of an EMG classification experiment. 

In future research, we plan to construct an expanded model of the proposed NN. 
As the function $g_i(y)$, which determines the shape of the distribution, was only examined in relation to $S_\mathrm{U}$ in this study, 
future work will investigate other functions. 
Despite of the assumption of $S_\mathrm{U}$ distribution, 
EMG data are occasionally distributed like a different type of distribution such as $S_\mathrm{B}$; 
thus in such situation $S_\mathrm{U}$ distribution is used as an approximation. 
Although $S_\mathrm{U}$ distribution worked well even in such situation according to the classification accuracy, 
more detailed comparison with other types of function and development of the selection criteria for the distribution type are needed. 
Using a different type of function for each dimension 
will also enable the classification of multivariate biosignals, such as the combination of EMG and EEG. 
Furthermore, the learning algorithm will be improved in future work, 
and the training time will be shortened by contriving numerical calculations for the Hessian matrix. 
Complete discriminative learning for ${}^{(1)}{\bf W}^{(c)}$ and ${}^{(2)}{\bf W}^{(c)}$ will also be developed 
using backpropagation-based learning.

	\appendices
	\section{Positive definiteness of the Hessian matrix}
	This appendix shows that the Hessian matrix described in (\ref{w3}) is positive semi-definite. 
As described in (\ref{HessianComponent}), 
the $(h, l)$th element of the $(c, k)$th block of the matrix is given as
\begin{eqnarray}
\lefteqn{\sum^{N}_{n=1}\frac{\partial ^2 E_n}{\partial {}^{(3)}w_{h}^{(c)}\partial{}^{(3)}w_{l}^{(k)}}} \nonumber \\
&=& \sum^{N}_{n=1}{}^{(5)}O_{k}^{(n)}(\delta_{c, k} - {}^{(5)}O_{c}^{(n)}){}^{(4)}O_{c,h}^{(n)}{}^{(4)}O_{k,l}^{(n)}.
\end{eqnarray}
For simplification, we can consider just one term in the summation over $n$, 
because the sum of positive semi-definite matrices is also positive semi-definite. 

Consider an arbitrary vector $\bm{u} \in C \times H$ with elements $u_{c,h}$. 
Then, 
\begin{eqnarray}
\lefteqn{\bm{u}^{\rm T} {\bf H} \bm{u}} \nonumber \\
&=& \sum^{C}_{c,k}\sum^{H}_{h,l}u_{c,h}{}^{(5)}O_{k}(\delta_{c, k} - {}^{(5)}O_{c}){}^{(4)}O_{c,h}{}^{(4)}O_{k,l}u_{k,l} \nonumber \\
&=& \sum^{C}_{c,k}b_c{}^{(5)}O_{k}(\delta_{c, k} - {}^{(5)}O_{c})b_k \nonumber \\
&=& \sum^{C}_{c,k}b_c b_k {}^{(5)}O_{k}\delta_{c, k} - \sum^{C}_{c,k}b_c{}^{(5)}O_{c}b_k{}^{(5)}O_{k} \nonumber \\
&=& \sum^{C}_{k}b_k^2 {}^{(5)}O_{k} - (\sum^{C}_{k}b_k{}^{(5)}O_{k})^2, 
\end{eqnarray}
where 
\begin{eqnarray}
b_c = \sum^{H}_{h}u_{c,h}{}^{(4)}O_{c,h}, \\
b_k = \sum^{H}_{l}u_{k,l}{}^{(4)}O_{k,l}.
\end{eqnarray}
Here, ${}^{(5)}O_{k}$ is the posterior probability satisfying $0 \leq {}^{(5)}O_{k} \leq 1$ and $\sum_{k}{}^{(5)}O_{k}$ $= 1$.
Furthermore, the function $f(b_c) = b_c^2$ is a convex function. 
Hence, we can apply Jensen's inequality \cite{jensen1906fonctions} to give 
\begin{eqnarray}
\sum^{C}_{k}b_k^2 {}^{(5)}O_{k} &=& \sum^{C}_{k}f(b_k) {}^{(5)}O_{k} \nonumber \\
 &\geq& f(\sum^{C}_{k}b_k{}^{(5)}O_{k}) = (\sum^{C}_{k}b_k{}^{(5)}O_{k})^2.
\end{eqnarray}
Therefore, 
\begin{equation}
\bm{u}^{\rm T} {\bf H} \bm{u} \geq 0.
\end{equation}
From the definition of definiteness, the Hessian matrix ${\bf H}$ is positive semi-definite.

}

\EOD

\end{document}